\documentstyle[12pt,a41,psfig]{article}

\newcommand{\AmS}{{\protect\the\textfont2 
A\kern-.1667em\lower.5ex\hbox{M}\kern-.125emS}} 
\begin{document}

\begin{center}
{\LARGE\bf Recent progress on the $h^{q,\bar q}_1(x,Q^2)$ Distributions and the Nucleon Tensor Charges}

\vspace{1cm}
{J. Soffer}

\vspace*{1cm}
{\it Centre de Physique Th\'eorique, CNRS
Luminy\\Case 907, 13288 Marseille Cedex 9, France}\\

\vspace*{2cm}

\end{center}

\begin{abstract}
We recall the definitions and the basic properties of the
transversity distributions $h^{q,\bar q}_1(x,Q^2)$ and the corresponding
nucleon tensor charges $\delta q(Q^2)$. We briefly comment on different
estimates from several phenomenological models and on the future possible
measurements with the polarized $pp$ collider at RHIC-BNL. Recent works on
the $Q^2$-evolution of $h^{q,\bar q}_1(x,Q^2)$ are also discussed and their
implications on a very useful positivity bound.
\end{abstract}



\vspace{1mm}
\noindent

In high-energy processes, the nucleon structure is described by a
set of parton distributions, some of which are fairly well known and
best determined by means of Deep Inelastic Scattering (DIS). In particular 
unpolarized DIS yields the quark distributions $q(x)$, for
different flavors $q=u,d,s, etc...$, carrying the fraction $x$ of the nucleon
momentum. They are related to the forward nucleon matrix elements of the
corresponding {\it vector} quark currents $\bar q\gamma^{\mu}q$, and likewise
for antiquarks $\bar q(x)$. Similarly from longitudinaly polarized DIS, one extracts the
quark helicity distributions $\Delta q(x)=q_+(x)-q_-(x)$, where $q_+(x)$
and $q_-(x)$ are the quark distributions with helicity parallel and
antiparallel to the nucleon helicity. Clearly the spin-independent
quark distribution $q(x)$ is $q(x)=q_+(x)+q_-(x)$. We recall that
for each flavor, the {\it axial charge} is defined as the first
moment of $\Delta q(x) + \Delta\bar q(x)$ namely,

\begin{equation}
\Delta q= \int_{0}^{1} dx\left[ \Delta q(x) + \Delta \bar q
(x)\right]
\end{equation}
and in terms of the matrix elements of the {\it axial} quark current
$\bar q\gamma^{\mu}\gamma^5q$, it can be written in the form

\begin{equation}\label{Delta}
2\Delta qs^{\mu} = <p,s|\bar q\gamma^{\mu}\gamma^5 q|p,s>\ ,
\end{equation}
where $p$ is the nucleon four-momentum and $s_{\mu}$ its
polarization vector. In addition to $q(x)$ and $\Delta q(x)$, for
each quark flavor, there is another spin-dependent distribution for quarks,
called the transversity distribution $h_1^q(x)$ related to the
matrix elements of the {\it tensor} quark current $\bar
q\sigma^{\mu\nu}i\gamma^5q$. The $h^q_1$ distribution measures the
difference of the number of quarks with transverse polarization
parallel and antiparallel to the proton
transverse polarization and similarly $h_1^{\bar q}(x)$for antiquarks. One also defines the {\it tensor charge} as
the first moment
\begin{equation}
\delta q=\int_{0}^{1} dx\left[ h_1^q (x) - h_1^{\bar q}(x)\right]\ ,
\end{equation}
which receives only contributions from the valence quarks, since those from
sea quarks and antiquarks cancel each other due, to the charge conjugaison
properties of the tensor current.

The existence of $h_1^q(x)$ was first observed in a systematic study
of the Drell-Yan process with polarized beams \cite{RS} and some of
its relevant properties were discussed later in various
papers \cite{AM,CPR,JJ}. We recall that $q(x)$, $\Delta q(x)$ and
$h_1^q(x)$, which are of fundamental importance for our
understanding of the nucleon structure, are all leading-twist
distributions. Due to scaling violations, these quark distributions depend
also on the scale $Q$ and their $Q^2$-behavior is predicted by the QCD
evolution equations. They are different in the three cases and we will come
back later to this important question. On the experimental side, a vast
programme of measurements in unpolarized DIS has been undertaken for more
than twenty five years. It has yielded an accurate determination of the $x$
and $Q^2$-dependence of $q$ (and $\bar q$) for various flavors. The $ep$
collider HERA at DESY is now giving us access to a much broader kinematic
range for $x$ down to $10^{-4}$ or smaller and for $Q^2$ up to $5.10^4 GeV^2$ or
so. From several fixed-targets polarized DIS experiments operating presently at CERN, SLAC
and DESY, we also start learning about the different quark helicity
distributions $\Delta q(x,Q^2)$, in some rather limited $x$ and $Q^2$ ranges,
{\it i.e.} $0.005<x<0.7$ and $<Q^2>$ between $2$ and $10 GeV^2$. Concerning
$h^q_1(x,Q^2)$ (or $h^{\bar q}_1(x,Q^2)$), they are not simply accessible in
DIS because they are in fact chiral-odd distributions, contrarely to
$q(x,Q^2)$ and $\Delta q(x,Q^2)$ which are chiral-even~\cite{JJ}. They can be
best extracted from polarized Drell-Yan processes with two transversely
polarized proton beams. For lepton pair production $pp\to\ell^+\ell^- X$
$(\ell = e, \mu)$ mediated by a virtual photon $\gamma^{\star}$, the double
transverse-spin asymmetry $A^{\gamma^{\star}}_{TT}$ reads

\begin{equation}
A^{\gamma^{\star}}_{TT} = \widehat a_{TT} \frac{\sum_{q}^{} e^2_q h^q_1
(x_a,M^2) h^{\bar q}_1 (x_b, M^2) + (a \leftrightarrow b)}{\sum_{q}^{} e^2_q q
(x_a,M^2) \bar q (x_b, M^2) + (a \leftrightarrow b)},
\end{equation}
where $\widehat a_{TT}$ is the partonic asymmetry calculable in perturbative
QCD and $M$ is the dilepton mass. The rapidity $y$
of the dilepton is $y=x_a-x_b$, and for $y=0$ one has $x_a=x_b=M/\sqrt{s}$, where $\sqrt{s}$ is the
center-of-mass energy of the $pp$ collision. Note that this is a
leading-order expression, which can be used to get a first estimate of
$A^{\gamma\star}_{TT}$ from different theoretical results for $h^q_1$ and
$h^{\bar q}_1$. If the lepton pair is mediated by a $Z$ gauge boson, one has
a similar expression for $A^Z_{TT}$ \cite{BS},namely
\begin{equation}
A^{Z}_{TT} = \frac{\sum_{q}^{} (b^2_q -a^2_q) h^q_1
(x_a,M^2_Z) h^{\bar q}_1 (x_b, M^2_Z)+ (a \leftrightarrow b)}{\sum_{q}^{} (b^2_q +a^2_q)q
(x_a,M^2_Z) \bar q (x_b, M^2_Z)+ (a \leftrightarrow b)},
\end{equation}
where $a_q$ and $b_q$ are the vector and axial couplings of the flavor q to the $Z$. However in the case of
$W^{\pm}$ production one expects $A^W_{TT}=0$, because the $W$ gauge boson is
a pure left-handed object ({\it i.e.}, $a_q=b_q$), which does not allow a left-right interference
effect associated to the existence of $h^{q,\bar q}_1$ \cite{BS}.
 Such experiments
will be undertaken with the polarized $pp$ collider at RHIC-BNL~\cite{M}, but
so far, we have no direct experimental information on the shape, magnitude and
$Q^2$-evolution of these quark and antiquark transversity distributions. This is badely
needed considering the fact that several theoretical models give rather
different predictions for the transversity distributions. For example the MIT
bag model~\cite{JJ} leads to $h^u_1(x)$, which is small for $x$ near zero 
and has a maximum value of $\sim 1.8$ for $x\sim 0.4$. This is in contrast to
the QCD sum rules calculations~\cite{IK}, which predict a rather flat behavior
for $h^u_1(x)$ around the value $0.6$ for $0.2\leq x \leq 0.5$. Let us also
mention the chiral chromodielectric model~\cite{BCD} which assumes for
simplicity that $h^q_1(x,Q^2_0)\simeq \Delta q(x, Q^2_0)$ for a very small
scale $Q^2_0$, {\it e.g.} $Q^2_0=0.16 GeV^2$. In this case the shape of
$h^u_1(x) $ is similar to that of the MIT bag with a larger maximum value of
$3.8$ or so for $x\sim 0.3$, as shown in Fig.1. For the $d$ quark, $h_1^d(x)$ is negative and smaller in magnitude, following the trend of the corresponding helicity distribution $\Delta d(x)$. Similarly we expect all the antiquark transversity distributions $h_1^{\bar q}$ to be one order of magnitude smaller (see for example Fig.4). The isovector contributions of $h^q_1$ and
$h^{\bar q}_1$ have been also calculated in the SU(3) chiral quark-soliton
model~\cite{KPG}.

\begin{figure}[ht]

    \centerline{\psfig{figure=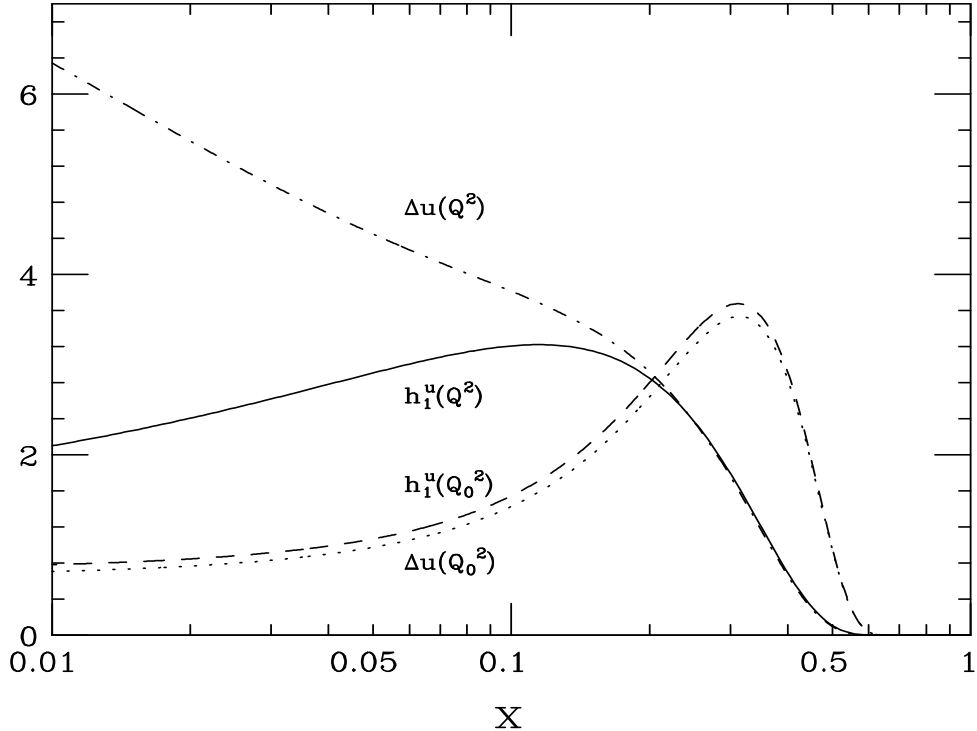,width=18cm,height=12cm,angle=90}}
    \caption[]{\small The $u$ quark helicity and transversity distributions $\Delta u $ and $h_1^u$, versus $x$ at the input scale $Q_0^2=0.16 GeV^2$ and evolved up to $25GeV^2$ (taken from ref.[8]). }
    \label{zpleptocc} 

\end{figure}

Concerning the axial charges and the tensor charges defined above, there are
various numerical estimates. In the non-relativistic quark model, they must
be equal as a consequence of rotational invariance. For example by using the
SU(6) proton wave function one finds,

\begin{equation}
\Delta u = \delta u = 4/3,~ ~ \Delta d = \delta d =
-1/3, \quad\hbox{and}\quad  \Delta s = \delta s = 0.
\end{equation}

So in this case the sum of the spin quarks (and antiquarks) is equal to the
proton spin {\it at rest} since we have
\begin{equation}
\Delta \Sigma \equiv \Delta u +\Delta d + \Delta s=1,
\end{equation}
but we get a wrong value for the axial-vector coupling $g_A=\Delta u-\Delta d
= 5/3$. Of course in polarized DIS, one is probing the proton spin in the
infinite momentum frame and the above result is surely no longer true. One
can evaluate the relativistic effects by making use of the Melosh
rotation~\cite{Me} and one finds for the axial charges~\cite{MMZ,BrSc},
\begin{equation}
\Delta u =1,\  \Delta d = -1/4 \quad\hbox{and}\quad \Delta s=0.
\end{equation}
In this case $g_A$ becomes $5/4$, in very good agreement with the
experimental value and $\Delta\Sigma$ gets also reduced from $1$ to $3/4$.
Although this shift goes in the right direction, this value is still too
large compared to the data, $\Delta\Sigma \sim 0.3$ or so, and it is very
likely that the discrepancy is due to a large contribution from polarized
gluons.

The effects of the Melosh rotation on the tensor charges have been calculated
in ref.~\cite{SS} and lead to
\begin{equation}
\delta u =7/6 \quad\hbox{and}\quad \delta d=-7/24,
\end{equation}
in remarkable agreement with the values obtained in the MIT bag
model~\cite{HJ}. However in ref.~\cite{KPG} they obtain
\begin{equation}
\delta u =1.12\quad\hbox{and}\quad \delta d=-0.42,
\end{equation}
but the large $N_c$ behavior is expected to generate in this model, large
theoretical uncertainties, mainly for the $d$ quark.
\begin{figure}[ht]

    \centerline{\psfig{figure=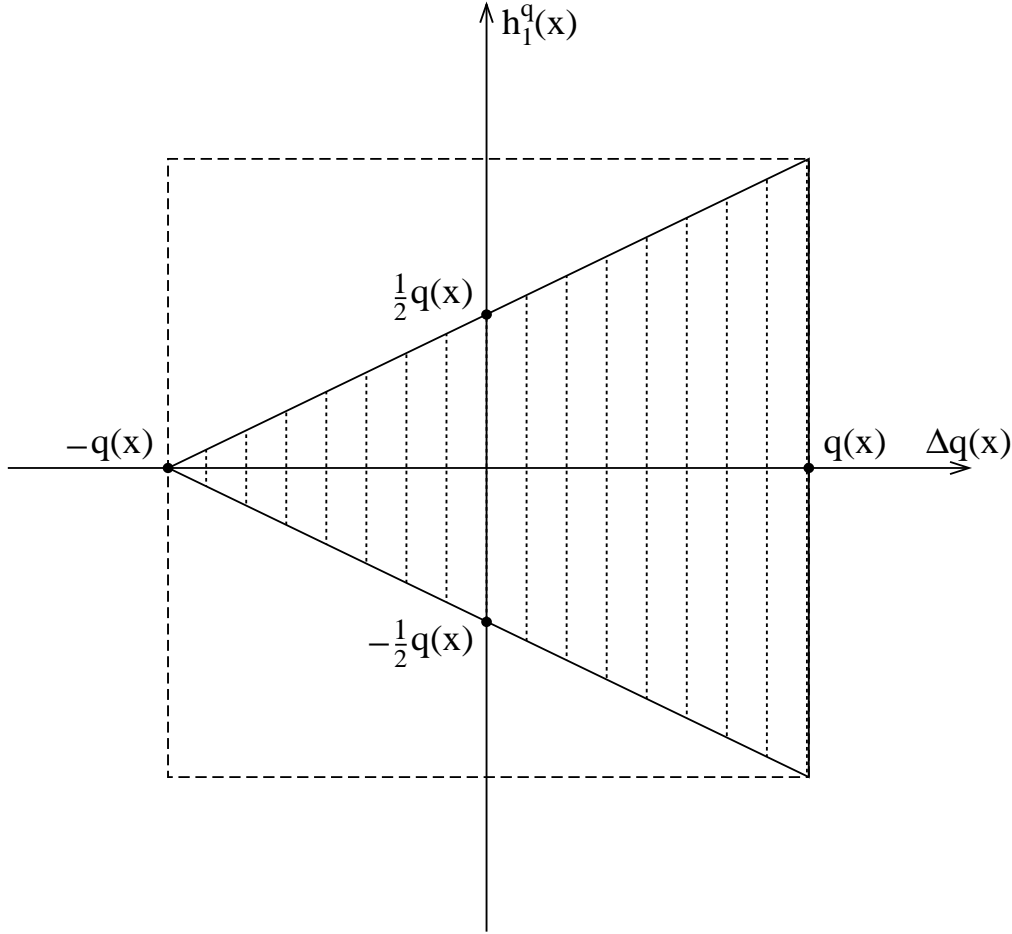,width=14cm,height=14cm}}
    \caption[]{\small The striped area represents the domain allowed by positivity (see eq.(12)).}
    \label{zpleptocc} 

\end{figure}

Now let us turn to a model-independent result. If we consider quark-nucleon
scattering, it can be shown that in the parton model $q(x),\ \Delta q(x)$ and
$h_1^q(x)$ are simply related to the imaginary parts of the {\it three} helicity
amplitudes $\phi_1$, $\phi_2$ and $\phi_3$ which are the only ones to survive
in the forward direction. From the positivity constraints among the $Im
\phi_i (0)$'s $(i=1,2,3)$, one finds on the one hand the trivial bounds
\begin{equation}
q(x)\geq 0 \quad\hbox{and}\quad q(x)\geq |\Delta q(x)|,
\end{equation}
and on the other hand, the following less obvious inequality~\cite{S}
\begin{equation}
q(x) + \Delta q(x) \geq 2|h^q_1(x)|.
\end{equation}
Clearly eq.(12) is more restrictive than the rather trivial bound which has
been proposed in ref.\cite{JJ} similar to eq.(11), namely
\begin{equation}
q(x)\geq |h^q_1(x)|,
\end{equation}
which does not involve $\Delta q(x)$.We show in Fig.2, the region allowed by eq.(12) which is half the region obtained by assuming eq.(13) instead. Indeed, in the very special situation
where $\Delta q(x) = q(x)$, eqs.(12) and (13) coincide, but it is not generally the case.
\begin{figure}[ht]

    \centerline{\psfig{figure=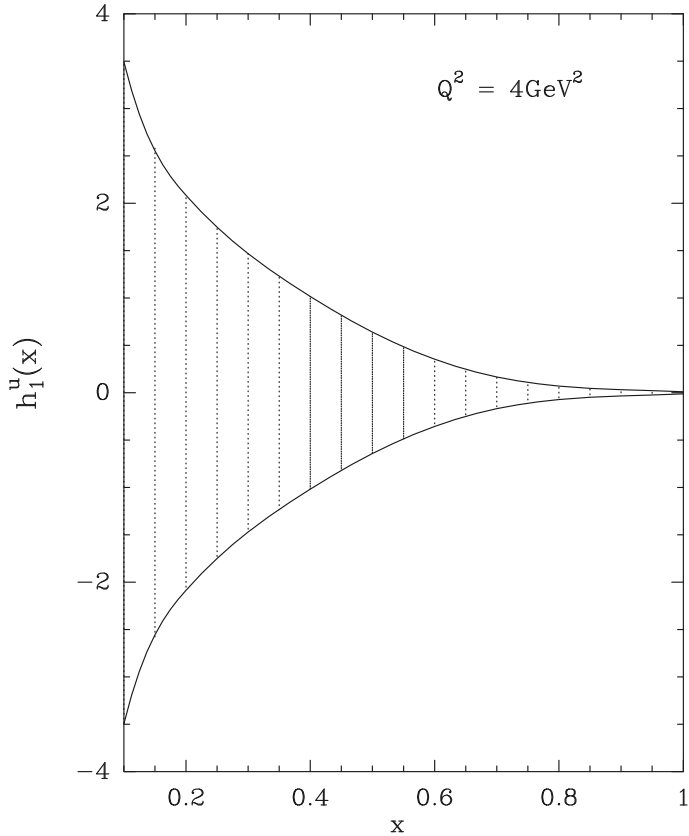,width=9cm,height=11cm}}
    \caption[]{\small The striped area represents the domain allowed for $h_1^u(x)$, using eq.(16) and ref.[16].}
    \label{zpleptocc} 

\end{figure}

 Needless to say that eq.(12) holds for all quark flavor
$q=u,d,s$ etc..., and as well as for their corresponding antiquarks. Obviously
any theoretical model should satisfy these constraints and we shall give some
examples. In a toy model \cite{AM} when the proton is composed of a quark and
a {\it scalar} diquark, one obtains the equality in eq.(12). In the MIT bag model,
let us recall that these three distributions are expressed in terms of two
quantities, namely one has \cite{JJ}
\begin{equation}
q(x) = f^2(x) + g^2(x),~ ~ \Delta q(x) = f^2(x) - 1/3 g^2(x) \quad\hbox{and}\quad h^q_1(x) = f^2(x) + 1/3 g^2(x),
\end{equation}
so in this case again the inequality (12) is saturated. To illustrate further
the practical use of eq.(12), let us assume, as an example, the simple
relation
\begin{equation}
\Delta u(x) = u(x) - d(x)
\end{equation}
proposed in \cite{BS1} and which is well supported by polarized DIS data. It is then
possible to obtain the allowed range of the values for $h^u_1(x)$ in terms of
unpolarized $u$ and $d$ quarks distributions since eq.(12) reads now
\begin{equation}
u(x) - 1/2 d(x) \geq |h^u_1(x)|.
\end{equation}

The allowed region is shown in Fig.3 and one can check, for example, that for $x\sim 0.4$ and $Q^2=4 GeV^2$ we get
$|h^u_1|\leq 1$, which must be obeyed by any phenomenological model.
\begin{figure}[ht]

    \centerline{\psfig{figure=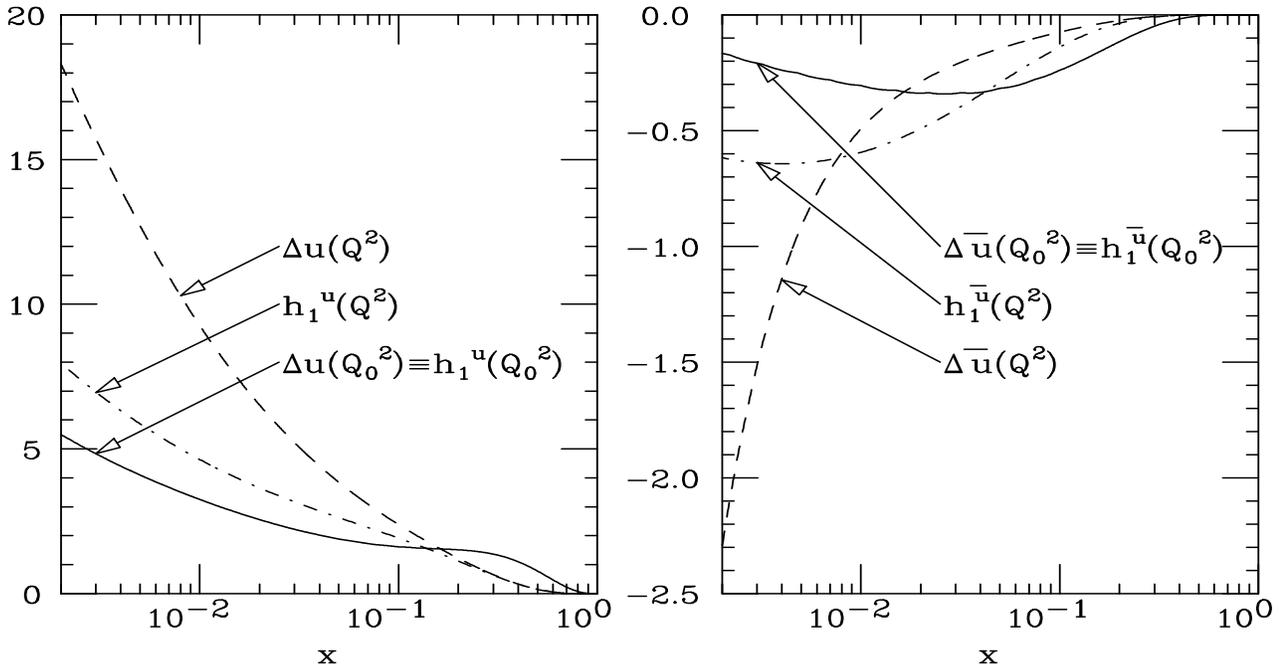,width=18cm,height=11cm,angle=90}}
    \caption[]{\small The $u$ quark helicity and transversity distributions $\Delta u $ and $h_1^u$, versus $x$ at the input scale $Q_0^2=0.23 GeV^2$ and evolved up to $25GeV^2$. The same for $\bar u$.(taken from ref.[8]).}
    \label{zpleptocc} 

\end{figure}

The positivity bound (12) has been rigorously proved in the parton model so
one may ask if it could be spoiled by QCD radiative corrections. Some doubts
have been expressed in ref.~\cite{GJJ}, where the authors claim that the
status of eq.(12) is similar to that of the Callan-Gross relation~\cite{CG}
which is known to be invalidated by QCD radiative corrections and becomes an
approximate equality at finite $Q^2$. We will come back to this objection, which will turn out to be not relevant, but meanwhile we want to discuss what is known about the $Q^2$
evolution of $h^{q,\bar q}_1(x,Q^2)$ and the corresponding tensor charge $\delta
q(Q^2)$. The Altarelli-Parisi equation for the QCD evolution of
$h^{q,\bar q}_1(x,Q^2)$ at order $\alpha_s$ is $(t\equiv log \ Q^2/\mu^2)$
\begin{equation}
\frac{dh^{q,\bar q}_1(x,t)}{dt} = \frac{\alpha_s(t)}{2\pi}\int^{1}_{x}
\frac{dz}{z} P_h(z) h^{q,\bar q}_1(x/z, t)
\end{equation}
where the leading order (LO) splitting function $P_h(z)$, which has been
obtained in ref.~\cite{AM}, reads
\begin{equation}
P_h(z) = \frac{4}{3}\left[ \frac{2}{(1-z)_+} - 2 + \frac{3}{2}
\delta(z-1)\right]=P^{(0)}_{qq}(z) - \frac{4}{3} (z-1).
\end{equation}
Here $P^{(0)}_{qq}(z)$ denotes the unpolarized LO quark-to-quark splitting
function calculated in ref.~\cite{AP} which is also equal to the
longitudinally polarized LO splitting function $\Delta_L P^{(0)}_{qq}(z)$ due
to helicity conservation. As a consequence of eq.(17) we see that $\Delta
q(x,Q^2)$ and $h^q_1(x,Q^2)$ have different $Q^2$ behaviors. In particular, if
for a given input scale $Q^2_0$ we have $\Delta q(x,Q^2)\simeq h^q_1(x,Q^2)$
after some evolution to $Q^2>Q^2_0$, one finds that mainly for $x<0.1$,
$h^q_1(x,Q^2)$ rises less rapidly than $\Delta q(x,Q^2)$, as shown for example
in Fig.1. This is a general property and in Fig.4, we show for illustration, the difference in the $Q^2$ evolution between $\Delta u ,\Delta \bar u$ and $h_1^u , h_1^{\bar u}$, for another set of distributions.
\begin{figure}[ht]

    \centerline{\psfig{figure=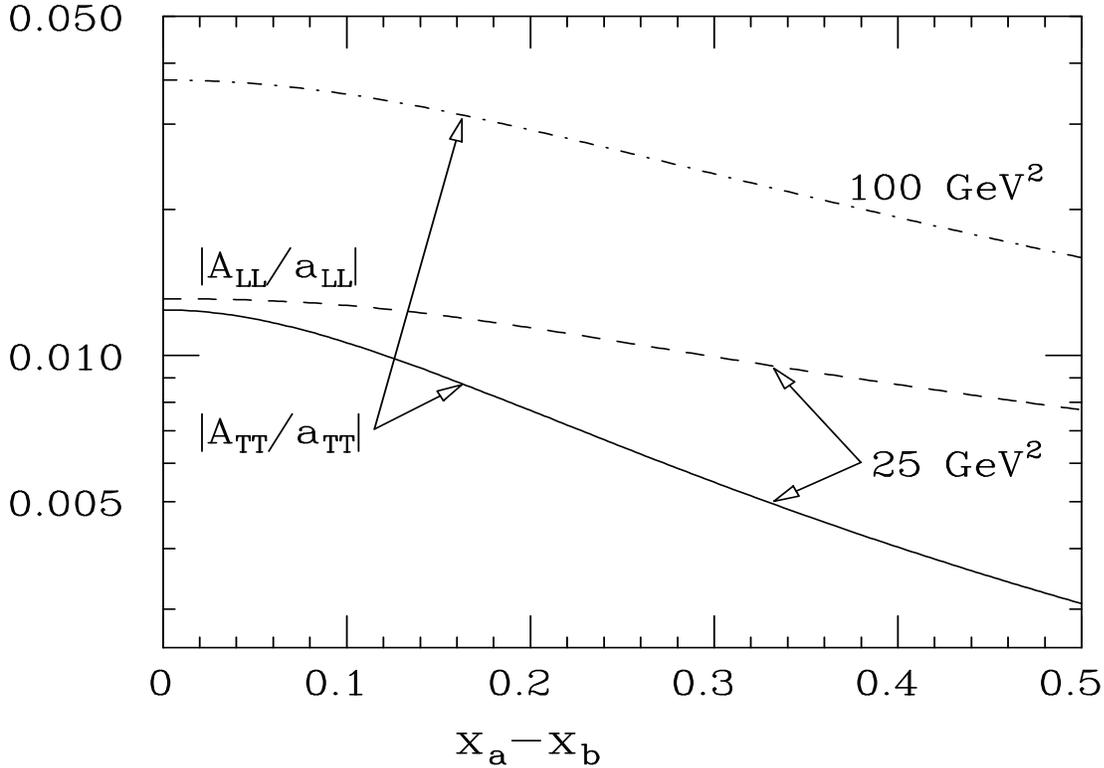,width=18cm,height=12cm,angle=90}}
    \caption[]{\small The Drell-Yan double transverse-spin asymmetry $|A_{TT}/a_{TT}|$ (see eq.(4)) for $pp$ collisions at $\sqrt s =100GeV$, as a function of $x_a-x_b$ (Solid line $M^2=25GeV^2$ and dot-dashed line $M^2=100GeV^2$). For comparison the double helicity asymmetry $|A_{LL}/a_{LL}|$ is shown for $M^2=25GeV^2$ (Dashed line).(taken from ref.[8]).}    \label{zpleptocc} 

\end{figure}

 A further consequence of eq.(18) is the $Q^2$ dependence of
the moments of $h^q_1(x,Q^2)$ and in particular the tensor charge which is
driven by the anomalous dimension $\gamma^h_1=-2/3$. Actually one finds that,
unlike the axial charge $\Delta q(Q^2)$ which remains constant, the tensor
charge $\delta q(Q^2)$ decreases with $Q^2$ since we have
\begin{equation}
\delta q(Q^2) = \delta
q(Q^2_0)\left[\frac{\alpha_s(Q^2_0)}{\alpha_s(Q^2)}\right]^{-4/27}.
\end{equation}
If one assumes as in ref.~\cite{BCD} that at $Q^2_0=0.16 GeV^2$ one has the
input tensor charges given by eq.(5), one gets at $Q^2=25 GeV^2$
\begin{equation}
\delta u = 0.969\quad\hbox{and}\quad \delta d = -0.25.
\end{equation}
The next-to-leading order (NLO) evolution of $h^q_1(x,Q^2)$ has been obtained
in three very recent papers~\cite{V,HKK,KM}. The results of these two-loops
calculations agree and show that, at NLO the tensor charge decreases with
increasing $Q^2$ even faster that at LO (see Fig.9 in ref.~\cite{HKK}).
 
Let us now come back to the $Q^2$ evolution of the inequality eq.(12). In a recent paper~\cite{B}, it was argued, by using eq.(18), that a sufficient condition to insure the validity of eq.(12) at $Q^2 >Q^2_0$,  if it is valid at $Q^2_0$, is that
\begin{equation}
\frac{|h_1^q|}{dt}<\frac{q_+}{dt}~,
\end{equation}
where $q_+=1/2[q+\Delta q]$. Strictly speaking the argument fails because $P_h(z)$ is {\it not} definite positive, but in a recent work~\cite{BST}, by means of a general mathematical method, it was shown that from the LO and NLO $Q^2$ evolutions, if the positivity
bound eq.(12) holds at a given $Q^2_0$, it is preserved at any $Q^2 >
Q^2_0$. The same conclusion was reached in ref.~\cite{MSSV}, using a numerical method.

 Finally some estimates can be made for the double
transverse asymmetry in dilepton production (see eq.(4)). Clearly at fixed
energy, $A^{\gamma^{\star}}_{TT}/\widehat a_{TT}$ increases with increasing
dilepton mass $M$, as shown in Fig.5, where we see that at RHIC energies, it will be at most $4\%$ for $\sqrt{s}=100GeV$ and
$M\sim 10GeV$. These predictions are confirmed in ref.[25], also in the case of the
$Z$ production and this small size is due to the small magnitude assumed for $h_1^{\bar u}$. Larger estimates ($\sim 10\%$ or so) have been obtained in
ref.~\cite{BS2}, but of course one must wait for the polarized $pp$ collider
at RHIC-BNL to be turned on by year $2000$.

It is my pleasure to thank the scientific organizers, J. Bl\"umlein and W.D. Nowak, for setting up this excellent workshop in such a
pleasant and stimulating atmosphere and P. S\"oding, for warm hospitality at DESY
Zeuthen.

\end{document}